\documentclass{aa-ext}  

\usepackage{natbib}
\bibpunct{(}{)}{;}{a}{}{,} 

\usepackage{graphicx}
\usepackage{color, xcolor} 
\usepackage{orcidlink}     
\usepackage{txfonts}
\usepackage{amsmath}
\begin{document}

  \title{Milliarcsecond astrometric oscillations in active galactic nuclei as a precursor of multi-messenger gravitational wave events}

   \author{L.I.~Gurvits\inst{1,2,3}\orcidlink{0000-0002-0694-2459}\fnmsep\thanks{leonid@gurvits.org}
          \and
          A.G. Polnarev\inst{4}\orcidlink{0009-0009-8870-2982}
          \and 
          S.~Frey\inst{5,6,7}\orcidlink{0000-0003-3079-1889}
          \and
          O.~Titov \inst{8}\orcidlink{0000-0003-1751-676X}
          \and
          A.A.~Osetrova \inst{9}\orcidlink{0009-0004-8160-1701}
          \and
          X. Fan 
          \inst{10}\orcidlink{0000-0003-3310-0131}
          \and
          A.~Melnikov \inst{9}\orcidlink{0000-0002-8466-7026}
          }

   \institute{Joint Institute for VLBI ERIC (JIVE), Oude Hoogeveensedijk 4, 7991 PD Dwingeloo, The Netherlands
         \and
             Faculty of Aerospace Engineering, Delft University of Technology, Kluyverweg 1, 2629 HS Delft, The Netherlands
        \and
            Shanghai Astronomical Observatory, Chinese Academy of Sciences, 80 Nandan Rd., Shanghai 200030, China
        \and
            Queen Mary University of London, London E1 4NS, United Kingdom 
        \and
            Konkoly Observatory, HUN-REN Research Centre for Astronomy and Earth Sciences, Konkoly Thege Mikl\'os \'ut 15-17, H-1121 Budapest, Hungary
        \and
            CSFK, MTA Centre of Excellence, Konkoly Thege Mikl\'os \'ut 15-17, H-1121 Budapest, Hungary
       \and
            Institute of Physics and Astronomy, ELTE E\"otv\"os Lor\'and University, P\'azm\'any P\'eter s\'et\'any 1/A,
            H-1117 Budapest, Hungary
        \and
            Geoscience Australia, PO Box 378, Canberra 2601, Australia
        \and 
             Institute of Applied Astronomy of the Russia Academy of Sciences, Kutuzova quay, 10, Saint-Petersburg, 191187, Russia 
        \and
            Steward Observatory, University of Arizona, 933 North Cherry Avenue, Tucson, AZ 85721, USA
             }

   \date{Received: 24 May 2024 / Accepted: 1 July 2025}
 
  \abstract
    {The existence of supermassive black hole binaries (SMBHBs) is predicted by various cosmological and evolutionary scenarios for active galactic nuclei. These objects are considered as contributors into the gravitational wave (GW) background, as well as emitters of discrete GW bursts. Yet, SMBHBs remain a rather elusive class of extragalactic objects. }
   {Previously we have identified the quasar J2102$+$6015 as a potential SMBHB system on the basis of absolute astrometric very long baseline interferometry (VLBI) monitoring. Here we present another case, the source J0204$+$1514, exhibiting a similar oscillating astrometric pattern. Our aim is to analyse the evolution of SMBHBs as generators of GW and provide a physical `multi-messenger' link between astrometric manifestation in the radio domain and GW emission.}
   {We analysed the available archive VLBI astrometry data that resulted in the detection of astrometric oscillations in the source J0204$+$1514. We assume these oscillations to be manifestations of orbital motion in a binary system. We estimated the parameters of the suspected SMBHB in this source and applied basic theoretical models to project its evolution towards coalescence. We also developed a simplified 'toy' model of SMBHBs consistent with the discovered astrometric oscillations and made quantitative predictions of GW emission of such sources using the case of J0204$+$1514 as an example potentially applicable to other SMBHBs.}
   {We provide observational evidence of astrometric oscillations in the source J0204$+$1514. As an ad hoc result, we also provide a re-assessed estimate of the redshift of J2102$+$6015, $z=1.42$. A toy model of the object containing a SMBHB with parameters consistent with the observed astrometric oscillations of the source J0204$+$1514 as an example enables us to consider GW emission as the cause of the system's orbital evolution.}
   {We conclude that astrometric VLBI monitoring has an appreciable potential for future detections of SMBHBs that could become multi-messenger targets for both electromagnetic (in radio domain) and GW astronomy. To outline the contours of a future physical model connecting SMBHB evolution with detectable GW manifestations, we present a toy model and, as an example, apply this toy model to the astrometrically oscillating source J0204$+$1514 described in this work. We also provide a justification for aiming future space-borne VLBI missions at direct imaging of SMBHBs as a synergistic contribution into future multi-messenger studies involving prospective GW facilities.}


\authorrunning{L.I.~Gurvits et al.}
\titlerunning{Milliarcsecond oscillations and multi-messenger studies}

  \maketitle
%

\section{Introduction}
\label{s:intro}

The formation of supermassive black hole binaries (SMBHBs) is one of the most challenging  problems of theoretical astrophysics and observational electromagnetic and gravitational wave (GW) astronomy. Their formation and subsequent evolution could be related to the merger of galaxies \citep{Begelman+1980Nat}, which might contain supermassive black holes (SMBHs) of any origin in their central regions. Mergers of `ordinary' black holes formed through stellar or galactic evolution may lead to the formation of SMBHBs.The latter scenario has been analysed for one of the first discovered black hole binaries in the core of the object 3C~66 \citep{DePaolis-2004-3C66}. Alternatively, black hole binary systems might be formed from primordial black holes (PBHs) born in the very early Universe \citep[e.g.][and references therein]{Polnarev+2012JCAP,Marcoccia+2023PBHB,PBH-merge-2024arXiv,escriva+2024PBH}.

However, regardless of the black hole binary genesis, they evolve and become `violent' at the last stages of their orbital evolution (sometimes referred to as recoiling or inspiralling) before coalescing and emitting more and more appreciable GWs. Recently, predictions about the SMBHB population were re-assessed by \citet{D'Orazio+2024book} and \cite{OVRO-SMBHB-2024arXiv}. On the basis of sinusoidal flux density variations detected in two blazars, the latter work predicts that about 1\% of blazars should contain SMBHB candidates.

A pivotal direct detection of GW was made by the LIGO and Virgo collaborations in 2015 \citep{GW-detect-2015PhysRevLett}. Although this discovery dealt with an act of coalescence of stellar-mass black holes, it stimulated an outburst of interest in the evolution of SMBHBs as generators of GW. The next powerful stimulus of interest in SMBHBs came recently with the announcement of the detection of a GW background by multiple pulsar timing array (PTA) monitoring programs, in particular by the  North American Nanohertz Observatory for Gravitational Waves, NANOgrav \citep[][and references therein]{NANOgrav-2023ApJ}. Very recent data from the \textit{James Webb Space Telescope (JWST)} in comparison with the NANOGrav GW background measurements apparently favour scenarios in which the SMBHBs are either `stalled' in their evolution or undergo accelerated inspiralling in the presence of gas in host galaxies \citep{SMBHB-JWST-2024}. New quantitative evaluations of SMBHBs' contribution to the GW background are reported in the recent data releases by PTAs \cite[][and references therein]{Goncharov+2024arXiv, Sato-Polito+2025arXiv}. It should be noted, however, that other potential sources of the stochastic GW background observed by PTAs may exist besides SMBHB coalescence \citep[e.g.][and references therein]{Braginsky+1990NCim, winkler2024origin, ferreira2024collapsing}.

According to \citet{Begelman+1980Nat} \citep[see also][and references therein]{Colpi-2014,DeRosa+2019}, the formation and evolution of SMBHB encompasses three phases: the pairing phase, the hardening phase, and the final coalescence phase. The inhabitants of the SMBHB `zoo', albeit small so far in numbers, are classified according to the primary evidence of the binary morphology, such as transiting, apparent, telescopic, and astrometric binaries \citep[see][section 6.2 and Table~3]{Ayzenberg+2023LRR}. While the number of SMBHB candidates and corresponding publications have grown very rapidly over the past several years, SMBHBs as a class of extragalactic objects remain rather elusive. The enigmatic apparent deficit of SMBHBs has become especially topical in view of the recent NANOGrav results \citep{Sato-Polito+2023}. Some optimism about finding more binary active galactic nuclei (AGNs) comes from the recent `varstrometric' \citep[i.e. variability-induced astrometric jitter in the photocenter of the unresolved system,][]{Hwang+2020ApJ} analysis of a \textit{Gaia}-based sample of quasars using Very Long Baseline Array (VLBA) observations \citep{Chen+Lazio+2023ApJ}. This study provides an indication that two out of 23 observed sources, J1044$+$2959 and J1110$+$3653, are possible binary quasars with the separation at the parsec scale. According to another rather optimistic prognosis, prospective millimetre/sub-millimetre VLBI studies (frequencies 86--690~GHz) will detect about 20 SMBHBs with masses of $M<10^{11}M_{\odot}$ at $z\le 0.5$ \citep{Zhao+2024ApJ}. A forward look at the contribution of LISA (Laser Interferometer Space Antenna) to studies of SMBHBs as multi-messenger sources in the context of their evolution through the hardening phase has recently been given by \cite{Spadaro+2025PhRvD}.

Arguably, one of the most `trusted' SMBHB candidates is the BL Lacertae object OJ\,287 \citep[][and references therein]{Britzen+2018MNRAS}. Its potential SMBHB nature has been corroborated by recent polarimetric space very long baseline interferometry (SVLBI) observations by the \textit{RadioAstron} mission at 22~GHz and nearly simultaneous Earth-based VLBI observations at 15, 43, and 86~GHz \citep{Gomez+2022ApJ}. In this source, indications of the binary nucleus come from observations of different types, including evidence of jet precession in VLBI images \citep{Britzen+2018MNRAS} and optical flares \citep{Dey+2018ApJ}. The estimates of the total mass of the binary system obtained with the VLBI and optical variability techniques show a notable discrepancy. The precession time scale analysed by \citet{Britzen+2018MNRAS} favours the total mass of the OJ\,287 system of $\sim 10^8 M_{\odot}$. The optical flare analysis  leads to the total mass of $\sim 10^{10} M_{\odot}$ \cite[][and references therein]{Dey+2018ApJ}. 

A somewhat similar case of a suspected binary system containing an SMBH is reported in the nucleus of a galaxy associated with the flare ASASSN-20qc on the basis of its `after-flare' quasi-periodic activity detected by multiple high-energy instruments \citep{pasham+2024periodic}. In this case, the authors demonstrate indications of a binary system containing an SMBH and an intermediate-mass black hole (IMBH).

Another recently reported suspected case of a sub-parsec-separated SMBHB is based on gravitational self-lensing manifestations detected in the Seyfert galaxy NGC\,1566 \citep{Self-lens-2024}. The more massive component in this binary system responsible for the lensing effect has an estimated mass of $5\times 10^{10}M_{\odot}$, with a secondary object ten times lighter. 

As another example, repeating major outbursts of the quasar J1048+7143 detected both in radio and $\gamma$-ray bands could be successfully modelled with the spin--orbit precession of the more massive jetted black hole with mass $M \approx 1.45 \times 10^9 M_{\odot}$, tightly constraining the ratio of the smaller black hole mass, $m$, to the mass of the larger BH: $0.062 < q=m/M < 0.088$ 
\citep{Kun+2022, Kun+2024}. 

To date, two SMBH components have been directly detected in only a few dual or binary systems. (Following the terminology of \citet{DeRosa+2019}, we define dual AGNs as not yet gravitationally bound accreting SMBH systems residing in the same host galaxy, and binary AGNs at sub-parsec projected separation.) One such system is the radio galaxy 0402$+$379 \citep{Rodriguez+2006,Bansal+2017}. In this source, monitored with multi-frequency VLBI imaging observations over 12~years, a pair of orbiting black holes with the total mass $M\approx 1.5\times10^{10}M_{\odot}$, identified in the milliarcsecond-scale morphology, demonstrates an orbital motion with the period $P\approx 3\times 10^{4}$~yr and a projected separation between the components of $\sim$7.3~pc. In a recent work, \citet{Surti+2024ApJ} give a somewhat higher value of the nucleus' mass in this object, $2.8^{+0.8}_{-0.8}\times10^{10}M_{\odot}$ (one of the heaviest black hole systems known), and argue that the pair of orbiting black holes in this source might already be gravitationally bound, and thus constitute an SMBHB.

A significantly larger apparent offset between two nuclei, $\sim 500$\,pc, has been reported in the recently identified and dynamically confirmed case of the SMBHB system NGC\,7277 \citep{Voggel+2022AA}. The authors note that the smaller SMBH in this system is in an advanced state of inspiral, and the mass ratio of the two SMBHs in this system is $q \approx 0.042$. A comparable separation between SMBHB components of $\approx 230$~pc is also reported in the object UGC~4211 on the basis of Atacama Large Millimeter/submillimeter Array (ALMA) observations \citep{Koss+2023ApJ}. A smaller separation of $\sim 100$~pc has recently been reported by \citet{Trinidad-Falcao+2024} in the resolved dual nuclei in the infrared galaxy MCG-03-34-64 at $z = 0.016$ on the basis of imaging and spectroscopic observations in radio (Very Large Array, VLA), optical (\textit{Hubble Space Telescope}), and X-ray (\textit{Chandra}, \textit{Suzaku}, and \textit{XMM-Newton}) bands. A new focused search for sub-kiloparsec dual AGNs based on HST and VLA imaging with several likely new candidates is presented by \cite{VODKA-2024}.

Several suspected dual AGN systems have been reported recently at kiloparsec scales. \citet{Xu+2024} investigated four such dual systems selected on the basis of SDSS and VLA observations. They emphasise that two systems with projected separations between dual AGNs of between 7 and 8~kpc contain components dubbed J0051$+$0020B and J2300$-$0005A with clear jet activity on parsec scales detected by VLBA. A new case of dual AGNs with a separation of $\sim$1~kpc and black hole mass ratio of $\sim$7:1 is reported by \cite{MaNGA-2025ApJ}. This dual system is located within an edge-on disc galaxy, SDSS J1445$+$4926, and consists of two BHs of masses $(9.4 \pm 2.7) \times 10^{6}M_\odot$ and $(1.3 \pm 0.26) \times 10^{6}M_\odot$. A case of two merging QSOs, HSC\,J121503.42$-$014858.7 and HSC\,J121503.55$-$014859 at one of the highest redshifts known to date for such systems, $z=6.05$, with a projected separation between the components of 12~kpc, was recently serendipitously discovered by the Subaru telescope \citep{Matsuoka+2024ApJ}. It should be noted that the population of apparent double or even triple quasars (AGNs) with separations of the order $1-100$~kpc is much more populous \citep[see, e.g.][]{Dual-AGNkpc-2022ApJ} than `likely' SMBHBs separated by parsecs and closer. However, a somewhat opposite trend has recently been reported by \cite{Saeedzadeh-2024}. They claim that the number of dual AGNs with separations of 0.5--4~kpc is twice the number of duals with separations of 4--30~kpc. This study also indicates an increase in the number of dual AGNs (as well as singles) from lower to higher redshifts.

To reveal the geometric and kinematic properties of SMBHBs, high-angular-resolution observational data are required. In this sense, VLBI comes as a natural technique in searching for and studying potential SMBHB systems \citep[e.g.][]{An+2018}, provided that the AGNs are radio-emitting. This technique offers two measuring products: source images and their astrometric parameters. One convincing example of SMBHB discovery using VLBI observations was reported for the radio galaxy 3C~66B \citep{Sudou+2003}. In that work, in-beam phase-referencing relative astrometry VLBI observations of the unresolved radio core in 3C~66B with respect to the background source 3C~66A resulted in the detection of the well-defined elliptical motion of the former with a period of $(1.05\pm 0.03)$~yr. A straightforward interpretation of such a kinematic pattern is Keplerian motion in a binary system. Several more cases of astrometry-based suspected SMBHBs are mentioned in \citet[][section 6.2.2 and references therein]{Ayzenberg+2023LRR}.

Recently we have found that several radio sources observed in VLBI astrometry monitoring programs demonstrate a remarkable `oscillating' behaviour in their celestial co-ordinates. Among others, these include J2102$+$6015 (2101$+$600) \citep{Titov+2023AJ, LIG+2023IAUS} and J0204$+$1514 (0202$+$149) presented in this work. We underline that these findings are based on absolute VLBI astrometric measurements and in this sense differ from the relative VLBI astrometric study of 3C~66B by \citet{Sudou+2003} mentioned above. We consider a Keplerian motion of the SMBHB's components as one of possible causes of the detected astrometric oscillations. Using this assumption, we further analyse their behaviour and evolution as potential contributors to the GW background and precursors of GW bursts. We note, however, that this is not a sole possible explanation of the apparent oscillations. The other plausible cause of the effect might be variations in the position of the centroid of radio brightness distribution due to consecutive emission of jet components and their motion. We note that this explanation requires a separate study as a follow-up on the work by \cite{Charlot-1990AJ} and in general is of high importance for improving the precision of the International Celestial Reference Frame \citep[ICRF,][]{Charlot+2020AA}.

In Section \ref{s:SMBHB-2cand}, we present the case of J0204$+$1415 as an astrometrically oscillating source. We briefly describe its properties known from the literature (subsection~\ref{ss:0202-ap-pro}) and analyse the VLBI astrometric measurements (subsection~\ref{ss:0202-anls}). The latter subsection contains a description of the statistical evaluation of the obtained parameters of the oscillations. Section~\ref{s:quali-eval} is dedicated to qualitative evaluation of the astrometric pattern of J0204$+$1415. In Section~\ref{s:toy}, we assume a toy model of the sources in which the detected astrometric oscillations are considered as manifestations of SMBHBs. We further analyse the orbital evolution of these SMBHBs and evaluate the GWs emitted by them, including the potential contribution of these or similar objects in the GW background detected by PTAs, as well as prospects of detecting GW bursts produced in final coalescence by SMBHBs. Section~\ref{s:Outlook} provides a brief outlook of a future search for SMBHBs as objects of multi-messenger studies. Section~\ref{s:conclu} lists several conclusions from the current study. In Appendix A, we present a new estimate of the redshift of the source J2102$+$6015, astrometric oscillations of which were reported by \cite{Titov+2023AJ}.

\section{Astrometric oscillations of the quasar J0204$+$1514}
\label{s:SMBHB-2cand}

The analysed datasets on J0204$+$1415 come from VLBI monitoring programs conducted in the framework of astrometric and geodetic studies by the International VLBI Service (IVS)\footnote{\url{https://ivscc.gsfc.nasa.gov}, accessed 2025.04.15}. The object of our interest has an appreciable radio emission of synchrotron nature (see subsection \ref{ss:0202-ap-pro} below). Therefore, it is natural to assume the existence of at least one accretion disc in the system.

The source was observed intensively by the geodetic IVS network for about 15 years until 2009. An apparently anomalous behaviour of J0204$+$1514 was reported in  the ICRF2 paper \citep{Fey+2015-ICRF2}, in which this source was referred to as a ‘specially handled’ target, i.e. one requiring a special procedure for obtaining astrometric solutions. This explains its exclusion from the regular IVS observational program after 2009. The available body of VLBI observations of this source enabled us to suspect that its astrometric parameters exhibit non-random behaviour.

\begin{figure*}[h]
    \includegraphics[width=\textwidth]{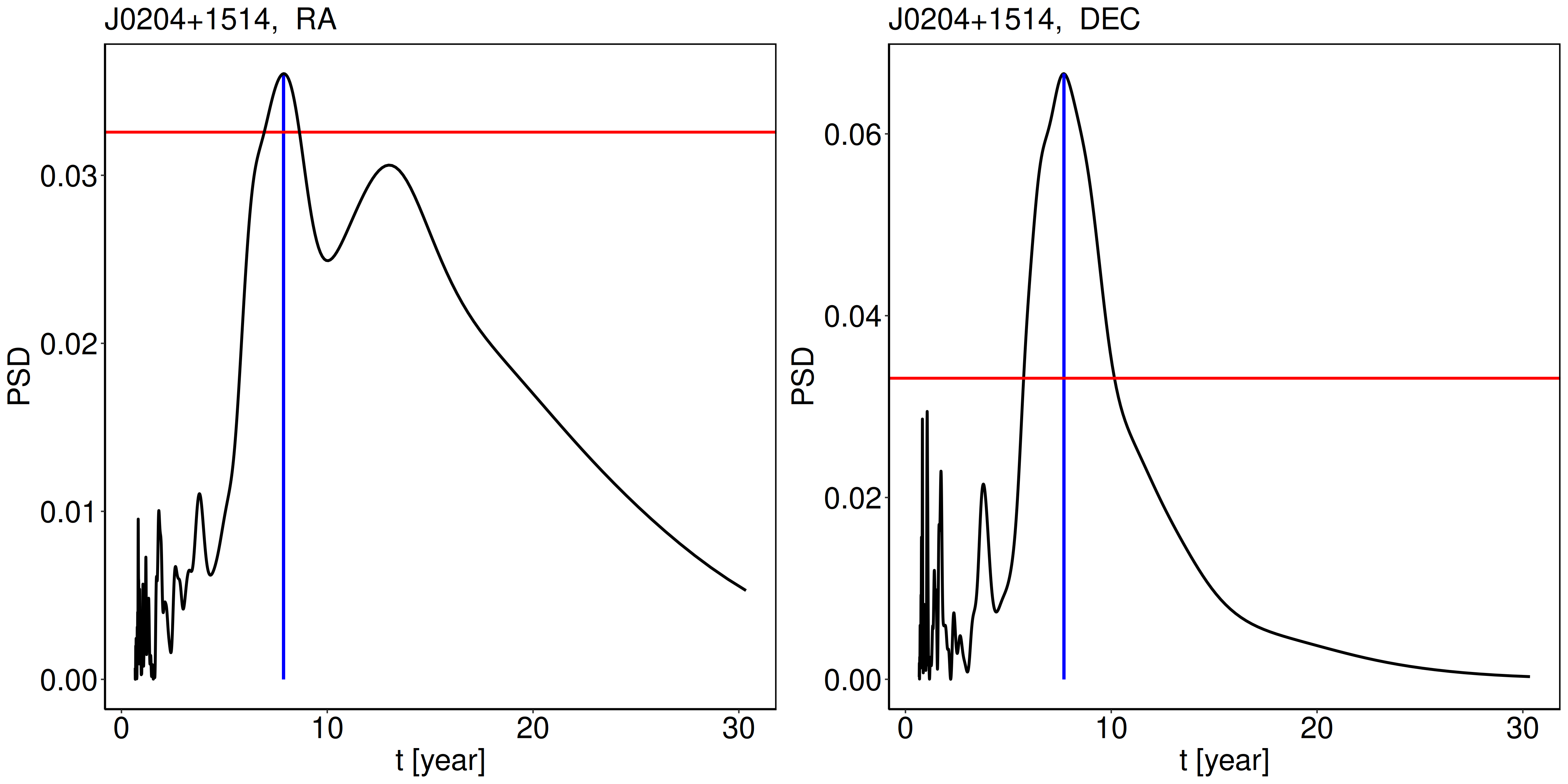}
    \includegraphics[width=\textwidth]{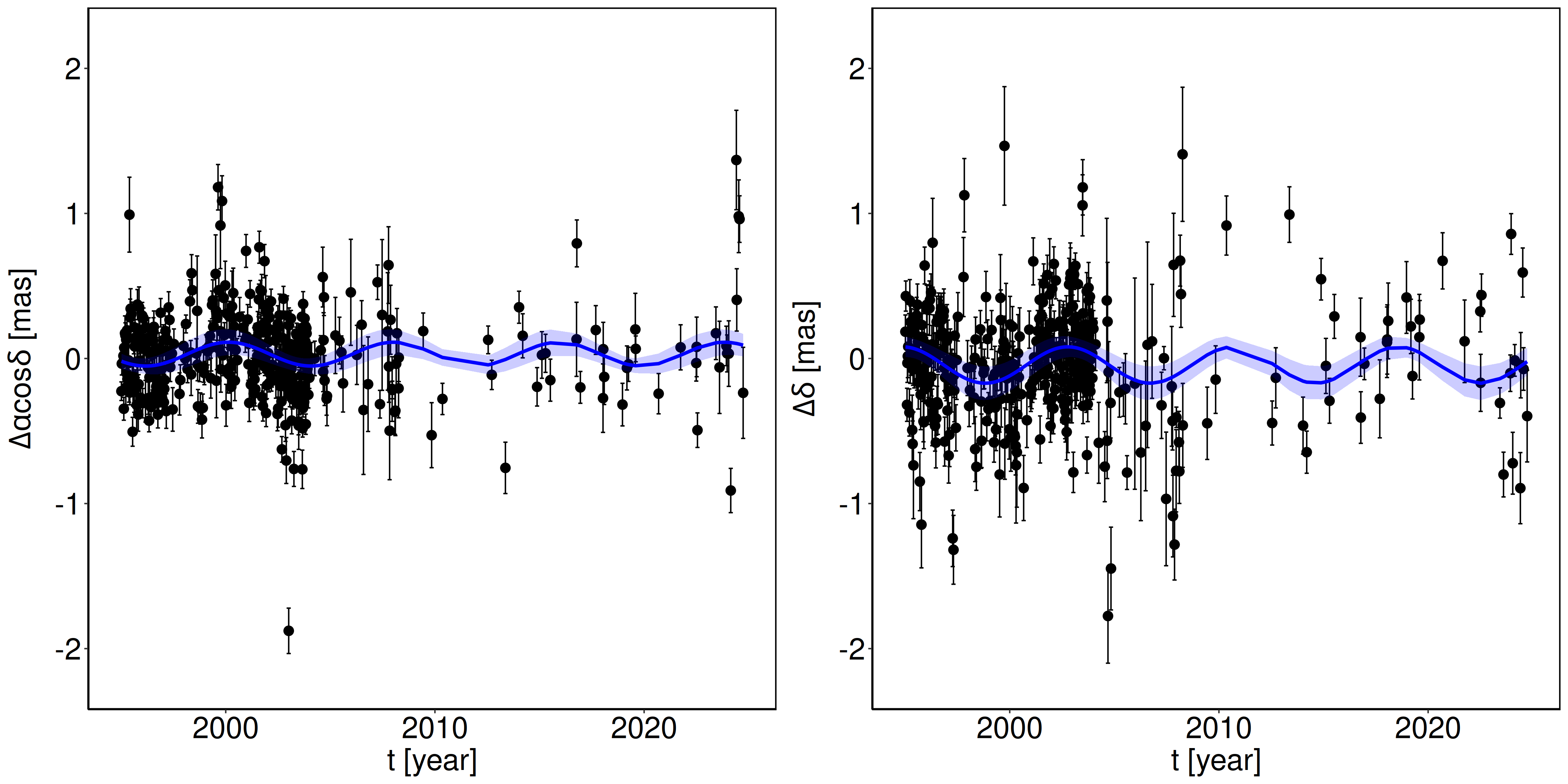}
    \caption{\textit{Top plots:} GLS spectra of RA (left) and DEC (right) of VLBI astrometry measurements of the source J0204$+$1415 over nearly 30 years.  The vertical blue lines indicate the peak (maximum) period value of 7.9 years in both co-ordinates. The horizontal red line marks the level below which the FAP (false alarm probability) is higher than 5\% \citep{VanderPlas-2018}. \\
    \textit{Bottom plots:} Temporal variations of RA (left) and DEC (right) of VLBI astrometric measurements centred to the J2000.0 position RA = $02^\mathrm{h}04^\mathrm{min}50.4138^\mathrm{s}$ and DEC = $+15\degr14\arcmin11.043\arcsec$ \citep{Xu+2019ApJS}. The trend with coefficients from Table~\ref{tab:1} has been removed. The observational data points are shown in black with their formal $\pm 1\sigma$ error bars. An apparent decrease in data time density after $\sim$2009 is due to the recognised anomalous astrometric behaviour and exclusion of the source from regular IVS observing runs (see Section~\ref{s:SMBHB-2cand} for explanation and reference). The blue curves represent the model fit with parameters listed in Table~\ref{tab:1}. The shaded blue area around the blue curves indicates the 99.73\% confidence interval ($\pm 3 \sigma$) for the model prediction. }
    \label{fig:0202-psd}  
\end{figure*}

\subsection{Brief review of astrophysical properties of the source J0204$+$1514}
\label{ss:0202-ap-pro}

The object J0204$+$1514 (PKS~0202$+$149, NRAO\,91, 4C$+$15.05) is a well-studied flat-spectrum variable radio source associated with an AGN (see a concise review of its properties by \cite{Jones_2018}). Its classification is uncertain. In some works, it is referred to as a blazar \citep{Piner+Kingham-1998ApJ}, while others classify it not too reliably as a Seyfert galaxy \citep{Veron-C+Veron-2006AA}. Its redshift has been recently estimated as $z= 0.834$ \citep{Jones_2018}. In the context of the current study, it is important to note that in many VLBI observations at centimetre wavelengths, the source demonstrates a core--jet morphology with apparent relative superluminal motion of its components with a speed of up to $15.9\,c$ \citep{Lister+2013AJ}, where $c$ denotes the speed of light. A hint of possible transverse motion of a jet component relative to the jet axis is reported on the basis of several VLBI observations by \cite{Piner+Kingham-1998ApJ}. The highest-resolution image of the source at 5~GHz obtained with the space VLBI \textit{VSOP} mission confirms the jet at sub-milliarcsecond scales \citep{Dodson+2008}. The source has an appreciable flux density at millimetre wavelengths \citep[e.g.][]{Owen+1981ApJ,WMAP-2009ApJ,Agudo+2014AA}. This, in combination with the astrometric oscillations discussed in the following subsection, makes the object J0204$+$1514 a suitable target for future multi-messenger studies discussed in Section~\ref{s:Outlook}.

\subsection{Analysis of astrometric VLBI data on J0204$+$1514}
\label{ss:0202-anls}

We applied several methods of analysing the observational VLBI astrometric data on the source J0204$+$1415 and defining the statistical confidence of the obtained models of astrometric variability.

\subsubsection{Generalised Lomb-Scargle periodogram}
\label{SSS:gls-p}

\begin{table*}[h!]
\caption{
        Amplitudes of terms for the models applied to the RA and DEC VLBI data.}
\begin{center}
    \begin{tabular}{|c||c|c|c||c|c|c|}
        \hline \hline
        & \multicolumn{3}{c||}{RA} & \multicolumn{3}{c|}{DEC} \\
        Term & Estimate & Phys. dimension & $p$ value & Estimate & Phys dimension & $p$ value \\
        (1) & (2) & (3) & (4) & (5) & (6) & (7) \\
        \hline
       constant & 21.4 $\pm$ 13.7 & $\mu$as & 0.12 & 148.7 $\pm$ 18.7 & $\mu$as & $8.8 \cdot 10^{-15}$ \\
        $(t-t_0)^3$ & --- & $\mu$as\,yr$^{-3}$ & --- & 0.2 $\pm$ 0.0 & $\mu$as\,yr$^{-3}$ & $2.6 \cdot 10^{-8}$ \\
        $(t-t_0)^2$ & 1.6 $\pm$ 0.2 & $\mu$as\,yr$^{-2}$ & $9.8 \cdot 10^{-14}$ & $-5.9 \pm 0.6$ & $\mu$as\,yr$^{-2}$ & $8.9 \cdot 10^{-21}$ \\
        $(t-t_0)$ & $-23.3 \pm 2.4$ & $\mu$as\,yr$^{-1}$ & $2.8 \cdot 10^{-20}$ & 19.8 $\pm$ 4.3 & $\mu$as\,yr$^{-1}$ & $4.7 \cdot 10^{-6}$ \\
        $\sin 2 \pi t /  7.9$ & $-73.7 \pm 15.6$ & $\mu$as & $2.9 \cdot 10^{-6}$ & 54.6 $\pm$ 21.2 & $\mu$as & 0.01 \\
        $\cos  2 \pi t / 7.9$ & $-17.3 \pm 16.1$ & $\mu$as & 0.3 & 147.9 $\pm$ 20.9 & $\mu$as & $3.4 \cdot 10^{-12}$ \\
        \hline
    \end{tabular}
    \tablefoot{The columns' contents:
    (1) -- the model terms; 
    (2) and (5) -- the estimated coefficients for RA and DEC, respectively; 
    (3) and (6) -- physical dimensions of corresponding coefficients; 
    (4) and (7) -- the corresponding FAP values. 
 The $p$ value indicates the probability of obtaining such a parameter from the observational data under the null hypothesis, i.e. assuming that the corresponding term has no effect. }
 \end{center}
    \label{tab:1}
\end{table*}

To quantitatively analyse the time series of VLBI astrometric measurements of the source J0204$+$1415, we first computed a generalised Lomb--Scargle (GLS) periodogram. The GLS method is widely employed for detecting periodicity in unevenly spaced data \citep{Lomb-1976ApSS, Scargle-1982ApJ}; see also \cite{VanderPlas-2018} for a comprehensive discussion of its practical applications in the analysis of experimental data. The technique involves fitting a periodic function, $\mu(f)$, using the (weighted) least-squares method and determining the optimal frequency, $f$, from a predefined frequency grid that minimises the residuals. In our analysis, we adopted the periodogram formalism developed by \citet{Baluev2014}.

Let us consider a time series of measurements, $x_i$, taken at times, $t_i$, each associated with an RMS uncertainty, $\sigma_i$. Each observation is thus represented as a triplet ($t_i$, $x_i$, $\sigma_i$), for which the periodicity (if any) is unknown. We consider two competing models for the data: the null (trend-only) model, denoted by $\mu_H(\theta_H, f, t)$, and the alternative model, given by
\begin{equation}
\mu_K(\theta_K, f, t) = \mu_H(\theta_H, f, t) + \mu(\theta, f, t),
\end{equation}
where
\begin{equation}
\mu(\theta, f, t) = A\sin  (2\pi f t)  + B \cos  (2\pi f t)
\end{equation}
is the periodic component, with $\theta = (A, B)$ and $\theta_H$ representing the parameter vector of the trend model. Clearly, the full parameter vector is $\theta_K = (\theta_H, \theta)$, and the frequency is defined as $f = T^{-1}$, where $T$ is the period.

The periodogram analysis allows us to estimate the power spectral density (PSD). The normalised PSD, $z(f)$, is given by the expression
\begin{equation}
z(f)  = \frac{\chi^2 _{H} - \chi^2_K(f)}{2},
\end{equation}
where
\begin{equation}
\chi^2 _{H, K}= \sum \frac{(x_i -\mu_{H, K}(t_i))^2}{\sigma_i^2}
\end{equation}
and $\chi^2_K(f)$ is calculated for each trial frequency, $f$, by fitting the model $\mu_K(\theta_K, f, t)$ to the data using the weighted least-squares method.

Several software packages are publicly available for computing the GLS periodogram; however, most of them fit either a purely periodic function or a periodic function with a constant term to the data, without accounting for linear or higher-order trends. As a result, it is often necessary to remove the trend from the data prior to computing the periodogram. Although this approach is not inherently incorrect, it may introduce a slight shift in the estimated period. Therefore, it is generally preferable to fit the periodic component simultaneously with the trend. We computed the periodogram using the GLS algorithm implemented in the Astropy library \citep{Astropy2013, Astropy2018}. Specifically, we used the astropy.timeseries.LombScargle class. 

In right ascension (RA), the quadratic trend was found to be significant, whereas the cubic term was not. In contrast, for declination (DEC), the cubic trend was significant. Therefore, we removed the corresponding trends before computing the periodogram: the quadratic trend
\begin{equation}
a (t - t_0)^2 + b (t-t_0) + c
\end{equation}
and the cubic trend 
\begin{equation}
a (t - t_0)^3 + b (t - t_0)^2 + c (t-t_0)
\end{equation}
before computing the periodogram. By default, the fitting model in astropy.timeseries is defined as
\begin{equation}
\mu_K(\theta_K, f, t) = \mu(\theta, f, t) + \mu_H(\theta_H, f, t) = \text{const} + \mu_H(\theta_H, f, t)  ,
\end{equation}
where a constant offset is included. Adding this constant term makes the approach more robust and statistically sound, especially in the presence of long-term trends in the data. The effectiveness of this method strongly depends on the form of the signal being modelled. If the true signal differs from the one we model, the fit may be inaccurate. In this analysis, we consider a harmonic signal with an additional trend component. However, one should keep in mind that the period may vary over time, and the actual signal shape may deviate from the assumed model.

\subsubsection{Statistical significance of the astrometric oscillations}
\label{sss:shnur-s}

The statistical significance of the highest peak in the PSD can be expressed in terms of the false alarm probability (FAP), which quantifies the likelihood of mistaking random noise for a genuine signal. One of the most reliable methods of estimating FAP employs the bootstrap approach \citep{Efron1979}. It evaluates the robustness of model fitting by repeatedly resampling the observed data with replacement data and refitting the model to each resample.  This procedure enables the assessment of the parameter uncertainty (FAP in our case) without relying on strong distributional assumptions. As is shown by \cite{VanderPlas-2018}, this approach might be more efficient than the Monte Carlo simulation in the context of PSD evaluation in the absence of reliable assumptions on the noise distribution. We generated a large number of synthetic datasets by randomly resampling the observed values with replacement data, while keeping the observation times fixed. For each resampled dataset, a GLS periodogram was computed, and the maximum peak recorded. The FAP was then estimated as the fraction of bootstrap iterations in which the maximum peak exceeds the peak observed in the original data.

Another approach involves the analytical approximation for the extreme value of FAP, developed by \cite{Baluev2008} using the theory of stochastic processes. This method is computationally efficient and particularly advantageous when analysing large numbers of time series, such as the ones encountered in large-scale astronomical surveys. It enables the fast and reliable estimation of the FAP associated with periodogram peaks, especially when spectral leakage is minimal or can be neglected. Even in the presence of significant aliasing, the approximation remains conservative and does not increase the likelihood of false detections.

To evaluate the statistical significance of the highest peak in PSD, we performed both bootstrap simulations and applied the \cite{Baluev2008} method to obtain FAP. Both of these methods are released in  astropy.timeseries.LombScargle. Our calculations provided the following values of FAP:
\begin{tabbing}
\= Right ascension: \= FAP$_{\text{bootstrap},\alpha} = 0.06$, \= FAP$_{\text{Baluev, RA}} = 0.02$ \kill
\> Right ascension: \> FAP$_{\text{bootstrap},\alpha} = 0.06$, \> FAP$_{\text{Baluev},\alpha} = 0.02$ \\
\> Declination: \> FAP$_{\text{bootstrap},\delta} = 0.0$, \> FAP$_{\text{Baluev},\delta} = 2.4 \times 10^{-13}$
.\end{tabbing}
These results suggest that the detected periodicity is statistically significant in DEC, but less so in RA.

Figure~\ref{fig:0202-psd} (upper panels, left and right) shows the GLS periodogram of the VLBI astrometry RA and DEC of J0204$+$1415, respectively. The highest significant peak in both RA and DEC occurs at the same period of $7.9$~years. The lower panels of Fig.~\ref{fig:0202-psd} show the periodic fit in the time series of VLBI astrometry of J0204$+$1415. Table~\ref{tab:1} summarises the best-fit amplitudes of trend components and harmonics together with their statistical confidence parameters, obtained in the analysis described above in subsection~\ref{SSS:gls-p}. 

\subsubsection{Validation of the periodicity by cross-spectrum}
\label{sss:val-n}

In order to validate the astrometric oscillation of the source J0204$+$1415 described above, we applied to the observational data the cross-spectrum method. It is used to identify periodic signals common to two time series by computing the Fourier transform of their cross-covariance function. The modulus of the cross-spectrum indicates the strength of shared periodic components; in our case, RA and DEC time series. This technique is particularly effective for detecting coherent periodicities that may be obscured in individual power spectra due to noise or aliasing \citep{Priestley1981}.

The autocorrelation function is related to the spectral density function through the Fourier transform. Specifically, the spectral density function, $g_{12}(\nu)$, is given by 
\begin{equation}
g_{12}(\nu) = \int_{-\infty}^{\infty} e^{-i 2\pi \nu \tau} \, \psi_{12}(\tau) \, d\tau  \,\,
\label{eq:cross_spectrum}
;\end{equation} 
\noindent hence,
\begin{equation}
 \psi_{12}(\nu) = \int_{-\infty}^{\infty} e^{i 2\pi \nu \tau} \, g_{12}(\tau) \, d\tau \,\, ,
\label{eq:cov_function}
\end{equation} 
where \( \psi_{12}(\tau) = <\epsilon_1(t) \epsilon_2(t + \tau)> \) is the cross-correlation function of two time series, \( g_{12}(\nu) \) is the corresponding cross-spectral power density, and \( \tau \) is the lag.

The cross-spectral density function, \( g_{12}(\nu) \), is generally a complex-valued function and can be represented in two equivalent forms:
\begin{equation}
g_{12}(\nu) = g_{12}^{+}(\nu) - i g_{12}^{-}(\nu),
\label{eq:cross_spec_real_imag}
\end{equation}
\begin{equation}
g_{12}(\nu) = \alpha_{12}(\nu) \, e^{i \phi_{12}(\nu)},
\label{eq:cross_spec_polar}
\end{equation}
where
\begin{itemize}
    \item \( g_{12}^{+}(\nu) \) and \( g_{12}^{-}(\nu) \) are the real and imaginary parts (or symmetric and antisymmetric components) of the complex function, \( g_{12}(\nu) \),
    \item \( \alpha_{12}(\nu) \) is the amplitude spectrum, and
    \item \( \phi_{12}(\nu) \) is the phase spectrum.
\end{itemize}

The amplitude spectrum is given by
\begin{equation}
\alpha_{12}(\nu) = \sqrt{ \left( g_{12}^{+}(\nu) \right)^2 + \left( g_{12}^{-}(\nu) \right)^2 }.
\label{eq:amplitude_spectrum}
\end{equation} 

\begin{center}
 \begin{figure}[!h]
    \includegraphics[width=0.48\textwidth]{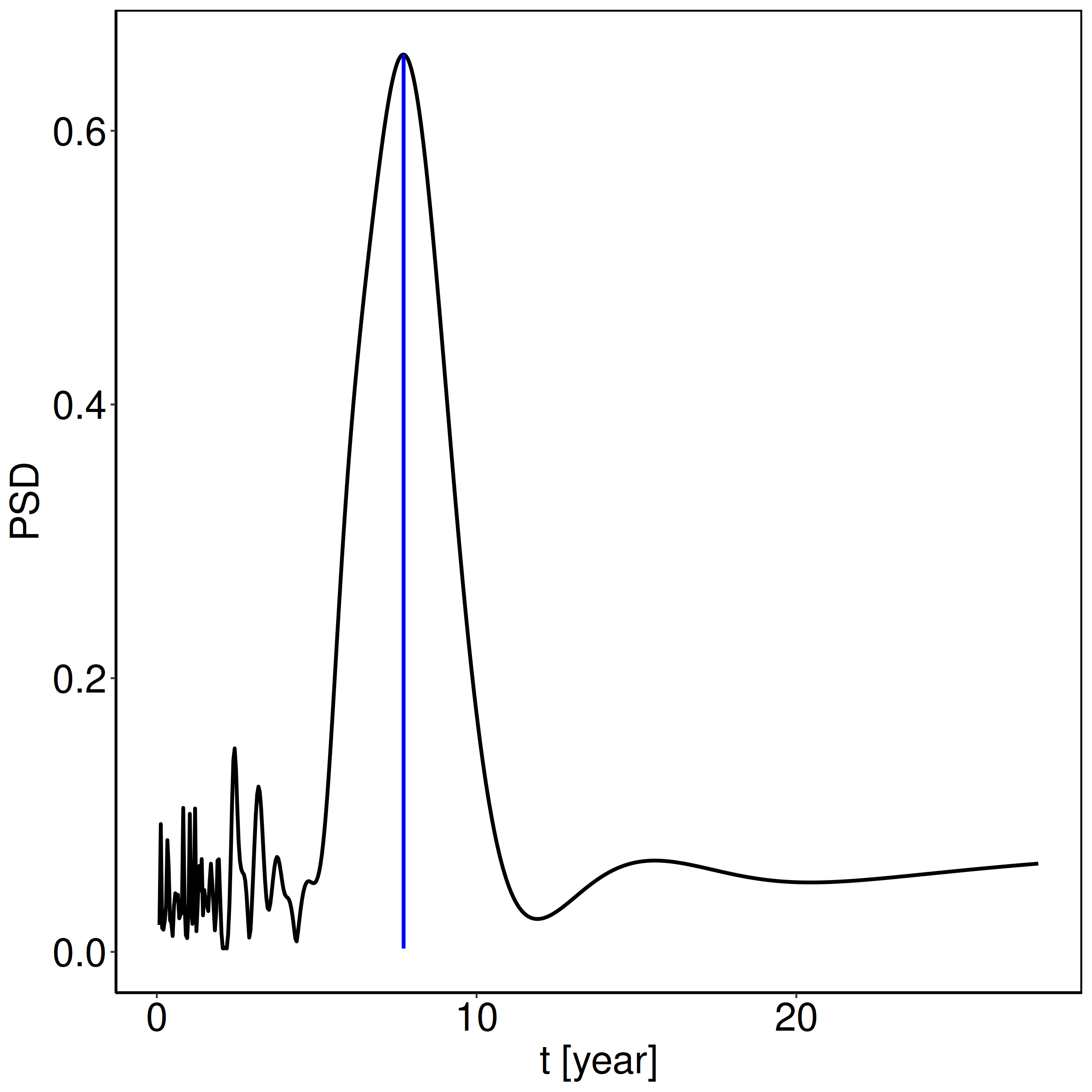}
    \caption{Cross-spectrum of RA and DEC VLBI measurements of the source J0204$+$1415. The peak is reached at the period $T = 7.8$~years.}
    \label{cross-specOLEG}
    \end{figure}
\end{center}

Figure~\ref{cross-specOLEG} shows the cross-spectrum, revealing a common period in RA and DEC of $7.8$~years. Although this value differs slightly from the result obtained with the GLS method in subsection~\ref{SSS:gls-p}, the difference appears to be small and could be a subject of a future detailed investigation. 

\section{Qualitative evaluation of the astrometric pattern of J0204$+$1514}
\label{s:quali-eval}

In the case of sources with complex milliarcsecond-scale structure, absolute astrometric positions could be affected by changes in the brightness distribution with time. In extreme cases, the effect of relative fading and brightening of components could cause an apparent `jump' up to $\sim 100$~mas in the radio source position \citep{Titov_2022}, \citep{Osetrova2024}. However, generating a persistent harmonic oscillating celestial trajectory this way is very unlikely.

The oscillating pattern of absolute celestial co-ordinate estimates for the source J0204$+$1514 might have a variety of explanations. We must note that VLBI astrometry deals with centroids of the source's brightness distribution, which are usually associated with the base of the jet in AGNs. These are optically thick regions of jets with typically flatter or inverted spectra at radio frequencies. The jet base areas do not coincide with the 3D position of central SMBHs. Thus, what is seen in Fig.~\ref{fig:0202-psd} is not the motion of the SMBHBs themselves, but their motion manifesting itself in the apparent motion of the astrometrically detected structural component of the source. 
A detailed physical description of the causal relationship between the binary system's orbital period and the observed period of astrometric oscillations associated with a jet-related morphological component warrants a dedicated investigation, which lies beyond the scope of the present paper. However, recent general relativistic magnetohydrodynamic (GRMHD) simulations of binary black hole systems suggest that, depending on the system's eccentricity, the periodicity of accretion and jet emission ranges from approximately 1.0 to 0.7 times the binary orbital period \citep{Manikantan+2025arXiv}. Therefore, in the subsequent sections, our assumption that the period of the SMBHB in J0204$+$1514 is equal to the observed period of astrometric oscillations introduces no significant error for the toy model considered.

Various effects might be invoked for explanations of the apparent oscillating behaviour of the jet base. One of the most studied cases, involving both long-term flux density variability monitoring in optics and radio, as well as structural studies at VLBI (milliarcsecond) scales, is the blazar OJ287, considered to be a likely SMBHB. However, in this and several other similar cases, alternative physical causes of temporal, structural, or astrometric manifestations, such as the impact of Lense--Thirring effect on an accretion disc or jet precession, are considered to be responsible for the observed phenomenology (for a review of the issue see \citet{Britzen+2023ApJ} and references therein). Among those causes, periodic quasi-orbital motion of the jet base associated with an orbital motion of SMBH in a binary system is at least no less plausible than others, especially coupled with the modality of SMBHB existence in the Universe (see Section~\ref{s:intro}). Moreover, SMBHBs must evolve, and the apparent orbital motion of an AGN, detected via oscillating behaviour of associated with bright morphological features in VLBI images of AGNs, could be looked at as an indicator of a potential presence of a SMBHB.

To a certain extent, our interpretation of an apparent oscillating astrometric pattern detected in the radio domain is similar to varstrometric detections of potential sub-kiloparsec dual AGNs based on \textit{Gaia} optical astrometry \citep{Hwang+2020ApJ}. Another search for potential SMBHBs via their quasi-periodic patterns in optical light curves was reported recently by \cite{Chen+Jiang-2024}.  We also note the recent work by \citet{Makarov+2024periodogram} in which a novel periodogram analysis algorithm applied to VLBI astrometric monitoring of 259 ICRF3 sources resulted in detection of 49 objects with quasi-periodic celestial position variations, including the source in which we are interested, J0204$+$1514. They report the detection of a quasi-periodic signal in DEC only, with a period of 8.05 years. This value slightly differs from our period estimate of 7.9 years. The reason for this discrepancy requires further investigation. We note, however, that, qualitatively, the absence of a detected quasi-periodic signal in RA for J0204$+$1514 by \cite{Makarov+2024periodogram} is consistent with the clearly lower confidence in periodicity detection in RA compared to DEC in our analysis (see Fig.~\ref{fig:0202-psd} and FAP values in Table~\ref{tab:1} and subsection~\ref{sss:shnur-s}).

Additional justification of our logic comes from the fact that if an SMBHB evolves all the way towards full coalescence, accompanied by a GW burst, it must end up as a single SMBH. Indeed, some known objects such as the quasar 3C~186 are suspected to be post-coalescence products of a former SMBHB \citep{boschini+2024}. Another suspected case of merging SMBHs or even a post-coalescence recoiling SMBH is reported recently in the radio quasar PKS~0903$-$57 based on optical spectroscopy \citep{Goldoni+2024AA}.

In the following section, we analyse the evolution of SMBHBs with astrometric oscillations. Such a binary system will eventually become a multi-messenger with very distinct electromagnetic and GW imprints. We do so for the source discussed in this work, J0204$+$1514. This source offers a straightforward example, allowing us to focus on properties of SMBHB as a GW generator. However, the analysis presented in the following section is applicable to other sources with apparent astrometric oscillations, and wider SMBHBs identified by other techniques. 

\section{Toy model of SMBHB evolution}
\label{s:toy}

As was mentioned in the introduction, the formation and evolution of SMBHBs encompasses three phases: the pairing phase, the hardening stage, and the final coalescence phase \citep[see][and references therein]{Begelman+1980Nat,Colpi-2014}. In this section, which is mainly devoted to the second (hardening) phase of the evolution of SMBHBs, we limit ourselves to an extremely simple (toy, as we call it) model, which is given solely in order to outline the contours of future detailed research (for a more detailed description of the evolution of the binary system at this phase, see e.g. \citet{D'Orazio+2024book}). This toy model consists of a set of simplifying assumptions that enable us to obtain a rough but helpful understanding of properties and GW-driven evolution of potential SMBHBs. As a numerical input into our toy model and its analysis, we use just one observed value of an astrometrically oscillating source, the period of the oscillation. This single value provides a link between the observational results described in previous sections of this paper and the analysis of astrometrically oscillating sources described below. Moreover, we elaborate the toy model using one source described in this work, J2102$+$6015, as a straightforward example. The approach and corresponding physical relations described below can be applied directly to any other SMBHB system.

\subsection{Formation and evolution of SMBHBs}
\label{ss:formation}

The first phase of SMBHB evolution is considered in a number of works \citep[e.g.][]{Kou+2024arXiv}, in which formation of SMBHBs is considered to be a result of collision of galaxies, in the nucleus of which there are single SMBHs. Based on the available statistical observational data, it is concluded that such binary systems are insufficient to explain the GW background detected using PTAs \citep[see, e.g.][and references therein]{NANOgrav-2023ApJ}. If we remain within the hypothesis that this GW background is generated by SMBHBs, we should assume that not all binary black hole systems are formed as a result of the merger of two galaxies. It is possible that only one of the two black holes, with the mass $M$, is surrounded by a dense medium in the central region of a galaxy, whereas the second, less massive black hole with the mass $m < M$ (it could be, for example, a PBH) is captured by a galaxy with the central BH of the mass $M$. Such events could be more frequent than collisions of galaxies and could generate the GW background mentioned above. SMBHBs still remain a rather elusive class of extragalactic objects not only from the observational perspective mentioned above, but also from the theoretical standpoint: presently, their abundance and mass function are very uncertain.

For the sake of simplicity, we assume in our toy model that the mass ratio of black holes in a binary system, $q=m/M$, is small. Recently, \citet{Kuznetsov-2024-OJ287} (for the object OJ\,287) and \citet{Titarchuk+Seifina-2024} (for the object SDSS~J075217.84+193542.2) demonstrated that X-ray observations
support this assumption and give $0.01 \lesssim q \lesssim 0.1$, which is also consistent with the estimate for the source J1048$+$7143 by \citealp{Kun+2024} (see Section~\ref{s:intro}).

Such an assumption makes it possible to describe the motion of a less massive black hole as the motion of a test body in a given field of a more massive central black hole. Then $a$ will be defined by Kepler's third law:
\begin{equation}
   a = \left(\frac{GM}{4\pi^{2}}\right)^{1/3}\left(\frac{T}{1+z}\right)^{2/3} \simeq 0.7\times 10^{-2}  M_9^{1/3}\left( \frac{T_3}{1+z} \right)^{2/3} \, \mathrm{pc},
\end{equation}
\noindent where $G$ is the gravitational constant, $M_9 =M/10^9 M_{\odot}$, $T_3 =T/3.0 \,\mathrm{yr}$, $T$ the observed period, and $z$ the redshift of the object of interest. For $M_9=1$, $T_{3}=2.63$, and the redshift of the source J0204$+$1415, $z=0.834$, we come to the following estimate of the separation between the binary system components:
\begin{equation}
   a \simeq 8.9 \times 10^{-3} \mathrm{pc} \simeq 89\, r_{g}. \label{eq:a}
\end{equation}
\noindent Such a value of separation, $a$, corresponds to an angular size of the order of 1~$\mu$as. This value is very close to the estimate obtained for the suggested SMBHB system in the object PKS~2131$-$021: the separation between the components $(0.001-0.01)$~pc with the rest-frame orbital period of $(2.082 \pm 0.003)$~yr \citep{O'Neill+2022ApJ}. However, this latter result is obtained by a method different from ours; namely, by radio flux density monitoring over more than 45 years. Recently, this result for PKS~2131$-$021 has been strengthened by a discovery with the same technique of a second, similar object, PKS~J0805$-$0111, with the rest frame period 1.422$\pm$0.005~yr \citep{OVRO-SMBHB-2024arXiv}. 

From Eq.~\ref{eq:a}, we can see that for the object J2102$+$6015, $a$ is considerably larger than $r_g$. This suggests the following simplification for our toy model: the relativistic corrections to all parameters of the secondary black hole orbit are not relevant for order-of-magnitude estimates. In particular, this means that we can consider the central black hole to be Schwarzschild rather than the Kerr one. In other words, we can neglect, along with all other relativistic effects, those related to the rotation of the central SMBH.

The next hardening phase, when the less massive black hole gradually gets closer to the central black hole, can naturally be divided into two distinct stages. During the first stage (considered in subsection~\ref{ss:fric-accr} below), the energy and angular momentum losses of already formed SMBHB  due to the radiation of GWs are negligible and the characteristic evolution time is determined by the interaction of the second, less massive black hole with matter around the more massive central black hole. During the second stage (considered in subsection~\ref{ss:GW-evoSMBHB}), when the two black holes are close enough to each other, the main losses of energy and angular momentum are due to the radiation of GWs. 

\subsection{Dynamical friction followed by interaction with the accretion disc}
\label{ss:fric-accr}

At the beginning of the first stage, the most significant role in the evolution of the SMBHB is played by dynamical friction, i.e. gravitational interaction with stars in the cluster around the central SMBH. As was shown in \citet{AGP+Rees-1994}, dynamical friction results in loss of energy and angular momentum by the secondary black hole in such a way that, in the most probable cases, the orbital eccentricity of the secondary black hole is rather small. For this reason (rather than only for simplicity), we assume that the eccentricity is zero, i.e. the orbit of the secondary black hole is assumed to be circular. 

This simplification means, in particular, that the astrometric oscillations noted above are not related to the Lidov--Kozai effect in celestial mechanics \citep{Lidov-1962P&SS,Kozai-1962AJ}. The application of this effect to SMBHBs deserves a separate investigation \citep[see, e.g.][]{Naoz-2016ARAA, Ivanov+2005MNRAS} and goes beyond the scope of this work.
 
The main problem associated with dynamical friction can be formulated as follows: whether the second black hole will manage in cosmological time (which is of the order $H_0^{-1}\sim 10^{10}$~yr, where $H_0$ is the Hubble constant) to get so close to the central black hole that dynamical friction ceases to be effective, but other factors come into play that lead to a faster convergence of black holes. Observational and theoretical solutions to this problem and its relation to the problem of the last parsecs are discussed in more detail in the review of \citet{D'Orazio+2024book}. We can make the following assumption, based on the conclusions made in this review:  for a wide range of parameters describing black holes and star clusters around the central black hole, the answer to this question is positive, i.e. we assume (without proof) that the timescale of reproaching of the secondary black hole to the central black hole due to dynamical friction in the cluster is considerably shorter than cosmological time. In other words, we assume that the problem of the last parsec is somehow solved. A more detailed analysis of the issue is beyond the scope of the present work.

As a less massive black hole approaches a more massive one, the role of the dynamic friction considered above in the evolution of the orbit gradually decreases. Sooner or later, the interaction of a less massive black hole with the accretion disc around the central black hole begins to play a major role in this evolution. A detailed analysis of this interaction is outside of the scope of the present work and must take into account that the accretion disc around the primary SMBH is not necessarily thin, and may even be strongly twisted, and that the orbit of the secondary black hole may be non-circular and arbitrarily oriented relative to the disc \citep[see, e.g.][]{Ivanov+Zhuravlev-2024MNRAS}. 

However, for our toy model it is possible to give a rough but sufficient estimate of the orbital evolution timescale of this interaction, $t_{\rm ev}$ \citep[see][]{Ivanov+1999MNRAS,Ivanov+2015AA}. If $m=q M$ is the mass of the smaller black hole,  and $\dot{M}$ is the accretion rate onto the more massive black hole, then
\begin{equation}
t_{\rm ev} \sim m/\dot{M} =qM/\dot{M}=10^7 q_{-2} M_9/\dot{M}_1 \,\, {\rm yr},
\label{eq:tev}
\end{equation}
\noindent where $q_{-2}=q/10^{-2}$ 
and 
$\dot{M_1}=\dot{M}/(1 \times M_{\odot}\,\, {\rm yr}^{-1})$ (see \citealp{Genzel+2024A&AR} for the motivation of such a normalisation of $\dot{M}$; see also \citealp{IMBH-mergers,D'Orazio+2024book}). According to Eq.~\ref{eq:tev}, for a reasonable choice of parameters, the timescale of the evolution of a SMBHB due to its interaction with the accretion disc is also less than the age of the Universe, just as the timescale of dynamical friction discussed above is.

\subsection{Gravitational wave emission and orbital evolution of SMBHB}
\label{ss:GW-evoSMBHB}

In this subsection, the characteristic timescale of evolution of the SMBHB  orbit due to the emission of GWs is compared to the corresponding characteristic timescale of orbit evolution defined by interaction with the accretion disc (see Eq.~\ref{eq:tev}). For a circular orbit and small $q$ \cite[see, e.g.][Chapter 13, Section 110]{Landafshitz-1975}, the separation between two components in a binary system changes with time according to
\begin{equation}
\dot{a} \approx \frac{2a^{2}}{GmM}\,\,\dot{E}_{\rm gw} \,,
\end{equation}
\noindent where $\dot{E}_{\rm gw}$ is the power carried away by GWs:
\begin{equation}
\dot{E}_{\rm gw} \approx -\frac{32G^{4}m^{2}M^{2}(m+M)}{5c^{5}a^{5}} \,\, .
\end{equation}
\noindent Hence, 
\begin{equation}
\dot{a} \approx -\frac{64 G^{3}mM(m+M)}{5c^{5}a^{3}} \approx -\frac{8cqr^{3}_{\rm g}}{5a^{3}} \,\, ,
\end{equation}
\noindent 
This  variation in $r$ corresponds to the timescale
\begin{equation}
t_{\rm gw} \sim -\frac{a}{\dot{a}} \approx \frac{5 r_{\rm g}}{8cq} \left(\frac{a}{r_{\rm g}}\right)^{4} \approx 2 \times 10^{-2}  (M_9/q_{-2})(a/r_{\rm g})^{4} \,\rm yr \, .
\label{eq:tgw}
\end{equation}

From $t_{\rm gw} \lesssim t_{\rm ev}$, and taking into account Eq.~\ref{eq:tev} and \ref{eq:tgw}, we find that the radiation of GWs plays a main role in the orbital evolution of SMBHB when 
\begin{equation}
a \lesssim r_{\rm gw} \approx 150\times q^{1/2}_{-2}\dot{M_1}^{-1/4}r_{\rm g} \, .
\label{eq:gwaves}
\end{equation}

As follows from Eq.~\ref{eq:a} and \ref{eq:gwaves}, we cannot exclude that the evolution of the source J0204$+$1415 is already determined by the radiation of GWs because $a < r_{\rm gw}$. The corresponding timescale of orbit evolution, as follows from (\ref{eq:a}) and (\ref{eq:tgw}), is
\begin{equation}
t_{\mathrm{gw}} \sim 2 \times 10^{-2}(M_9/q_{-2})\times (89)^{4}\,\mathrm{yr} \approx 1.2 \times 10^{6}(M_9/q_{-2}) \,\, \mathrm{yr}\,\, .
\label{eq:tgw-j2102}
\end{equation}
\noindent Hence, we can predict that, in approximately one million years (which is very soon by cosmological standards), this object is likely to become a source of detectable gravitational radiation.

Since the statistical distributions of major physical parameters (masses and mass ratios, spins, all orbital parameters) of the SMBHBs as a class of objects are largely unknown, we refrain from making any other predictions in this paper about the GW manifestations of SMBHBs for shorter time intervals. Nevertheless, in the next two subsections we present some considerations on the possibility of detecting GWs from individual SMBHBs by laser interferometer antennas such as \textit{LISA} \citep{LISA-Red-Book} (in subsection~\ref{ss:detectableGWs1}) and the possible contribution of SMBHBs to the GW background detected by PTAs \citep[][and references therein]{NANOgrav-2023ApJ} (in subsection~\ref{ss:detectableGWsbackground}). These considerations outline the contours of  future studies of SMBHBs as multi-messenger objects.

\subsection{Detection of gravitational waves generated by an individual SMBHB}
\label{ss:detectableGWs1}

In this subsection, we touch briefly on the future detection of GWs by  \textit{LISA}-type detectors and give some rough constraints on SMBHB parameters for which such detection is possible. Let us assume that a prospective GW detector can detect all GWs if (\textit{i}) their frequencies are in the range 
\begin{equation}
\omega_\mathrm{min} \lesssim \omega_{\mathrm{gw}} \lesssim \omega_\mathrm{max} \,\, ,
\label{eq:freqrange}
\end{equation}
and (\textit{ii}) their  amplitudes, $h$ (such an amplitude is called  the strain of a GW), are larger than some value, $h_0$ (called the sensitivity of the detector).

Such an assumption is a rough replacement for the non-monotonic dependence of the sensitivity of any real GW detector on the frequency of the receiving signal (see e.g. \citet{LISA-Red-Book} for \textit{LISA} and \citet{NANOgrav-2023ApJ} for NANOgrav). The main purpose of such an assumption is not only to simplify the restrictions given below on the parameters of SMBHBs, but also to make the physical meaning of these restrictions clearer.

When the SMBHB orbit is approximately circular, we have
\begin{equation}
\omega_{\rm gw} \approx 2\,\omega_{\rm orbit}(1+z)^{-1} \approx 2(GM)^{1/2}(1+z)^{-1}a^{-3/2} \,\,  
;\end{equation}
\noindent hence,
\begin{equation}
a \approx r_{\rm g}(c/r_{\rm g}\omega_{\rm gw})^{2/3}(1+z)^{-2/3} .
\end{equation}
We can see from Eq.~\ref{eq:tgw-j2102} that GW radiation can be detected only if
\begin{equation}
(c/r_{\mathrm{g}})^{2/3}\omega^{-2/3}_{\mathrm{max}}(1+z)^{-2/3} \lesssim a/r_{\mathrm{g}} \lesssim (c/r_{\rm g})^{2/3}\omega^{-2/3}_{\mathrm{min}}(1+z)^{-2/3} ,
\end{equation}
i.e.
\begin{equation}
[M_9\omega _{-4,\mathrm{max}}]^{-2/3} \lesssim a/r_{\mathrm{g}} \lesssim  [M_9\omega _{-4,\mathrm{min}}]^{-2/3} \,\, ,
\label{eq:arg1}
\end{equation}
where $\omega _{-4,\mathrm{max}}=10^4\omega_{\mathrm{max}}(1+z)$ and $\omega _{-4,\mathrm{min}}=10^4\omega_{\mathrm{min}}(1+z)$ \,\, . 

Another constraint comes from a comparison of the amplitude of the emitted GW with $h_{0}$. To make an order-of-magnitude estimate, we can use the quadrupole formula  \citep[see][Chapter 13, Section 110]{Landafshitz-1975}. Omitting tensor indices we get
\begin{equation}
h \approx \frac{2G}{3c^{4}R_L}\ddot{D} \,\,, 
\end{equation}
where $\ddot{D}\approx qMa^2\omega^2_{\mathrm{orbit}}$ is the
second time derivative of the quadrupole moment, $\omega_{\mathrm{orbit}}=(2GM/a^3)^{1/2}$ 
is the orbital angular velocity, and $R_L$ is the luminosity distance from 
the SMBHB to the detector of GWs. Then one can obtain the following inequality:
\begin{equation} 
h\approx \frac{qr^{2}_{\rm g}}{a R_L}\approx 10^{-15}(q_{-2}M_9/R_{1L})(a/r_{\rm g})^{-1}>h_{0},
\label{eq:h}
\end{equation} 
where $R_{1L} = R_L/1\,{\mathrm{Gpc}}$ \,\,.

From Eq.~\ref{eq:h}, we have another constraint on $a$:
\begin{equation}
a/r_{\rm g}\lesssim \frac{q_{-2} M_9}{R_{1L}h_{-15}} \,\, ,
\label{eq:arg2}
\end{equation}
where $h_{-15}=h_0\times 10^{15}$.

The last, very simple constraint on $r$ comes from the well-known fact that the radius of the last stable circular orbit around a Schwarzschild black hole is $3r_{\mathrm{g}}$. Hence, 

\begin{equation}
a/r_{\rm g}\gtrsim 3 \,\, .
\label{eq:lsco}
\end{equation}
Constraints on the parameters of SMBHBs can be obtained as conditions for the compatibility of inequalities \ref{eq:arg1}, \ref{eq:arg2}, and \ref{eq:lsco}:
\begin{equation}
\frac{q_{-2} M_9}{R_{1L}h_{-15}} \gtrsim (M_9\omega _{-4,\mathrm{max}})^{-2/3}
\end{equation}
and 
\begin{equation}
3 \lesssim (M_9\omega _{-4,\mathrm{min}})^{-2/3} \,\, .
\end{equation}

Finally, we have the following constraints on the parameters of SMBHBs generating detectable GWs \citep[see also][]{Ivanov+2015AA}:

\begin{equation}
 \frac{q_{-2} M_9^{5/3}}{R_{1L}}\gtrsim h_{-15} \omega_{-4,\mathrm{max}}^{-2/3} 
\label{eq:dist}
\end{equation}
and
\begin{equation}
 M_9\lesssim 0.2 \times \omega_{-4,\mathrm{min}}^{-1} \,\, . 
 \label{eq:mass}
\end{equation}

We note that combining inequalities \ref{eq:arg1}, \ref{eq:arg2}, and \ref{eq:lsco} leads to physically meaningful constraints on the masses and distance to the SMBHB in terms of the parameters of the GW detector (Eq.~\ref{eq:dist} and \ref{eq:mass}).

\subsection{SMBHBs and GW background detected by pulsar timing arrays}
\label{ss:detectableGWsbackground}

This subsection is devoted to the assumed population of SMBHBs that contributes to the GW background radiation detected by PTAs \citep[][and references therein]{NANOgrav-2023ApJ}. Let us assume that the population of SMBHBs is evolving with the characteristic time given by Eq.~\ref{eq:tev}, and consists of $N_x$ objects with approximately the same parameters. The number of SMBHBs, $N_*$,  that could  make a significant contribution to the GW background detected by PTAs above a given sensitivity threshold, $h_0$, and in a given frequency range (Eq.~\ref{eq:freqrange}) is $N_{*}=f N_x$, where $f \sim t_{\mathrm{gw}}/t_{\mathrm{ev}}$. As follows from Eq.~\ref{eq:tev} and \ref{eq:tgw},  
\begin{equation} 
f \approx 2\times 10^{-9} q_{-2}^{-2}\dot{M_1}(a/r_{\rm g})^{4} \,\, .
\label{eq:f}
\end{equation} 

The SMBHB separation, $a$, still satisfies inequalities \ref{eq:arg1} and \ref{eq:lsco}, 
while inequality \ref{eq:arg2} should be significantly modified. Indeed, for background radiation, inequality \ref{eq:h} should be rewritten as 
\begin{equation}
qr^{2}_{\rm g}/(a R_L) \gtrsim h_{0}/\sqrt{N_*}=h_0 (N_x f)^{-1/2}. 
\label{eq:nx}
\end{equation}
Hence, instead of Eq.~\ref{eq:arg2}, as follows from Eq.~\ref{eq:f}, we get
\begin{equation}
a/r_{\rm g} \gtrsim 2 \times 10^{4}\frac{h_{-15} R_{1L}}{M_9(N_x\dot{M_1})^{1/2}}.
\label{eq:arg3}
\end{equation}
The constraint on $N_x$ can now be obtained as a condition for the compatibility of inequalities \ref{eq:arg3} and \ref{eq:arg1}:
\begin{equation}
2 \times 10^{4}\frac{h_{-15} R_{1L}}{M_9(N_x\dot{M_1})^{1/2}}\lesssim  [M_9\omega _{-4,\mathrm{min}}]^{-2/3}.
\end{equation}
Hence,  
\begin{equation}
N_x \gtrsim 5 \times \frac{h^2_{-15} R^2_{1L}\omega _{-10,min}^{4/3}}{M_9^{2/3}\dot{M_1}},
\end{equation}
where $\omega _{-10,\mathrm{min}}=10^{10}\omega_{\mathrm{min}}(1+z)$.

It is notable that the parameter $q$ in our extremely simplified toy model does not appear in the above constraint for detectability of GWs by PTAs, contrary to the constraint in Eq.~\ref{eq:dist} for detectability of individual SMBHB GWs generated. The characteristic time of the existence of close-enough SMBHBs is proportional to $q$ (see Eq.~\ref{eq:tev}), while the characteristic time when individual SMBHBs contribute to the background radiation is inversely proportional to $q$ (see Eq.~\ref{eq:tgw}). Hence, when we deal with the GW background, the fraction of contributing SMBHBs is proportional to $q^{-2}$ (see Eq.~\ref{eq:f}). For this reason,  the corresponding requirement for the amplitude of the GW, which itself is proportional to $q$ (see Eq.~\ref{eq:arg2}), is weakened by a factor of $q$ (see Eq.~\ref{eq:f} and \ref{eq:nx}). The fact that the intensity of background GW radiation generated by SMBHBs does not depend on $q$ (or depends weakly in a model more accurate than the toy model presented in this paper) favours the formation of SMBHBs from less numerous but more  massive SMBHs and  more numerous  but less  massive SMBHs. 

\section{Outlook for future direct imaging studies of SMBHBs as multi-messengers }
\label{s:Outlook}

As was shown above, getting decisive information on the parameters of an SMBHB as a generator of GWs requires its measurable parameters to be obtained at the microarcsecond scale, at least indirectly. Arguably, the most convincing evidence and in-depth study opportunity would be provided by direct imaging observations of potential SMBHB objects with a microarcsecond resolution, in which the duality of the source's morphology would be visible directly, as in several cases mentioned in Section~\ref{s:intro} but at much smaller separations of milliparsecs -- see the estimates presented in Section~\ref{s:toy}. This can be illustrated following a very convenient visual representation presented in Fig.~1 by \citet{D'Orazio+2024book}: the dual AGN currently occupying the very top edge of the plot with the rest-frame periods of the order of $10^6$~yr should be `pushed down' by five orders of magnitude to orbital periods of the order of years in the area of VLBI astrometry. 

Microarcecond angular resolution is beyond reach of any existing astronomical imaging instrument or facility in any domain of the electromagnetic spectrum. The currently record-holding Earth-based VLBI instruments cannot provide such a resolution due to the fundamental limitations of the longest available baselines, which cannot exceed the Earth's diameter, and the Earth's atmosphere opacity at wavelengths shorter than about 0.7~mm. The only plausible solution for achieving the required angular resolution is to place a VLBI system in space, above the Earth's atmosphere and with no limitations on the baseline length. Several concepts of such systems have been developed over the past decade. One of them, THEZA (TeraHertz Exploration and Zooming-in for Astrophysics) has been presented in response to the European Space Agency's call for inputs into the prospective strategic program Voyage~2050 \citep{LIG+2021ExpA, LIG+2022Acta}. It suggests a multi-element space-borne VLBI configuration able to operate at millimetre and sub-millimetre wavelengths with an angular resolution of the order of 1~$\mu$as. Another concept called BHEX (Black Hole Explorer) considers an Earth--Space VLBI configuration operating at millimetre/sub-millimetre wavelengths \citep{BHEX-2024-SPIE}. With a bit of luck, and accepting the prediction by \citet[][section 3.1]{D'Orazio+2024book}, the currently ongoing analysis of the RadioAstron non-imaging AGN Survey \citep{Kovalev+2020ASR}, reaching in the best cases an angular resolution of $~10\,\,\mu\rm{as}$, might offer one to several new SMBHB candidates. However, they will still require decisive imaging observations with future space-borne VLBI systems. 

Another potentially efficient multi-messenger approach to investigating the contribution of mergers of black hole binaries into the GW background at frequencies higher than the ones explored by PTA, accessible to LISA, is suggested by \cite{Ellis+2024arXiv}. Their model reconciles the pre-JWST and JWST observations of the co-growth of masses of stellar population of galaxies and their central black holes, including SMBHBs. The quantitative estimates of GW generation by evolving SMBHBs presented in section~\ref{s:toy} might become useful in assessing the observational perspectives of LISA and other GW instruments covering specific frequency ranges.

While discussing prospects of future detections of SMBHBs as `visually' binary AGNs, we need to underline the difference between traditional morphological VLBI studies of jets in AGNs, especially their innermost bases, and the SMBHs themselves. The latter would become accessible for direct imaging observations if the technique demonstrated by the millimetre-wavelength VLBI Event Horizon Telescope (EHT) in the detection of the black hole shadows in M\,87$^*$ \citep{EHT-M87-I-2019} and Sgr~A$^*$ \citep{EHT-SgrA-I-2022} improved its angular resolution at least by an order of magnitude. Again, this would become possible only with prospective space-borne extensions of EHT such as BHEX and THEZA. 

\section{Summary and conclusions}
\label{s:conclu}

Astrometric VLBI monitoring with a sufficient angular resolution holds significant potential for advancing the study of SMBHBs during the hardening phase of their evolutionary pathway, and possibly even during the immediate pre-coalescence stage. The brief and by no means exhaustive overview of the SMBHB landscape provided in Section~\ref{s:intro} underscores the importance of each new SMBHB detection. Particularly valuable are those candidates identified in the late stages of their evolution, approaching coalescence, as these represent prime targets for multi-messenger observations. Although astrometric oscillations in AGNs cannot, in isolation, serve as conclusive evidence for the presence of an SMBHB, such systems remain among the most plausible explanations for the astrometric behaviour observed. We also briefly outline the prospect of directly imaging SMBHBs in advanced evolutionary stages as a compelling scientific objective for future space-based millimetre/sub-millimetre VLBI missions. We anticipate that the forthcoming detection of GWs by LISA, when combined with astrometric VLBI monitoring, will significantly advance our ability to confirm the existence and characterise the properties of SMBHBs.

   \begin{enumerate}
      \item The study presented in this paper is based on ad hoc peruse of VLBI astrometric measurements conducted for purposes other than a search for possible SMBHB systems. The major feature that attracted our attention was the apparent regular oscillations of the sky position of the source J2102$+$6015 (reported in our previous paper \cite{Titov+2023AJ}) and the new case of J0204$+$1514 presented in the current work) with sub-millarcsecond amplitudes. We focus our attention on the latter source as a case study and interpret the oscillating astrometric pattern as a manifestation of orbital motion in a SMBHB system with a period of $\sim 7.9$ yr in the observer's frame. We underline that neither of these sources is confirmed beyond doubt as containing SMBHBs. Nevertheless, a Keplerian motion of components in the suspected SMBHBs is a very plausible explanation of the observed astrometric oscillations. By stating this, we emphasise that our study does not and cannot rule out another reason for the apparent astrometric behaviour; for example, variations in the distribution of radio brightness in the sources due to their jet dynamics. A more careful re-analysis of abundant available astrometric VLBI data for other sources, including several dozen sources reported by \citet{Makarov+2024periodogram}, might provide indications of new SMBHB candidates. 

      \item Prompted by our interest in the potential presence of a SMBHB in the object J2102$+$6015 -- whose oscillatory behaviour was presented in our previous study \citep{Titov+2023AJ} -- a careful re-examination of the available optical spectrum has led us to propose in Appendix~\ref{s:J2102-red} a more plausible redshift value of $z = 1.42$, in place of the value previously adopted. This might be useful for future detailed studies of this source.
      
      \item The method of evaluating astrometric VLBI time series presented in the current study allowed us to detect a harmonic oscillating astrometric pattern for J0204$+$1514, with an apparent period of $\sim 7.9$\,yr. We emphasise that the goal of the present study is to describe quantitatively how observed astrometric oscillations of AGNs containing an SMBHB can be used to evaluate GW generation by such an object. This multi-messenger link between astrometric patterns detected in the radio domain and GWs is of fundamental value and does not depend on specific quantitative values of astrometric oscillations (periods).
      
      \item Assuming that the detected oscillating astrometric pattern is a manifestation of SMBHBs, we develop a toy model of such a system that is linked to the observationally detected astrometric oscillations with just one parameter, the oscillation period. The toy model of J0204$+$1514 allows us to investigate the evolution of orbital motion of a SMBHB and predict its manifestations as a multi-messenger source generating GWs. We quantitatively evaluate the potential detectability of GW emission generated by our source by the \textit{LISA} GW telescope and the contribution of a SMBHB with the parameters of our toy model to the GW background detected by PTAs.      
 \end{enumerate}

\begin{acknowledgements}
We are grateful to the anonymous referees for very insightful comments and useful suggestions, although, respectfully, we did not agree with some of them.
The authors gratefully acknowledge a very helpful advice on the presented here study by \u{Z}eljko Ivezi\'{c} and Ben Hudson. 

The authors acknowledge use of VLBI data obtained within the framework of the International VLBI Service for Geodesy and Astrometry (IVS). 

LIG gratefully acknowledges support by the International Talent Program of the Chinese Academy of Sciences, project number 2024PVA0008.

SF thanks the Hungarian National Research, Development and Innovation Office (NKFIH, grant no. OTKA K134213) for support. His work was also supported by the NKFIH excellence grant TKP2021-NKTA-64. 

\end{acknowledgements}

\bibliographystyle{aa}
\bibliography{SMBHB-lib}

\begin{appendix} 
\section{On the redshift of the source J2102$+$6015}
\label{s:J2102-red}

Our attention to possible astrometric oscillations of quasars was attracted by the source J2102$+$6015 as reported in \cite{Titov+2023AJ}. That work was dedicated to VLBI astrometric study of a sample of high redshift targets. Originally, the object J2102$+$6015 was identified as a blazar at the redshift of $z=4.575$ by \citet{Sowards-Emmerd+2004}. However, the authors labelled this identification as marginal. We note that in the context of the present study, the exact value of the redshift of this object is not critically important. Nevertheless, if numerical estimates described in Section~\ref{s:toy} are repeated for the source J2102$+$6015, a specific value of its redshift is important. Therefore, we have decided to revisit the evaluation of the optical spectrum of this source that has provided the estimate of redshift.

In the discovery paper by \citet{Sowards-Emmerd+2004}, a low dispersion spectrum of the object taken with the 9.2-m Hobby Eberly Telescope was presented. It shows a strong emission line at $\sim6800$\,\AA, which has been interpreted as a Ly$\alpha$ emission line. However, a more careful examination of the spectrum allows us to conclude that there is no strong Lyman break absorption bluewards of the Ly$\alpha$ emission line typical of high-redshift quasars. The usually strong \ion{C}{IV} 1549~\AA\, line was not detected either. Therefore, the emission line is not likely to be that of a high-redshift Ly$\alpha$. With only a single emission line in the spectrum, the exact redshift cannot be uniquely determined. However, due to the lack of other strong lines in the wavelength range covered by the observed spectrum, most likely this line corresponds to the \ion{Mg}{ii} 2800~\AA\ transition. According to this interpretation, the redshift of this source is $z=1.42$.

\end{appendix}

\end{document}